\documentclass[pdftex]{spie}
\usepackage[dvips]{graphicx}
\usepackage{amsmath}

\title{Electronic Dynamics Due to Exchange Interaction with Holes in Bulk GaAs}

\author{Hans Christian Schneider and Michael Krau\ss{}
\skiplinehalf
Physics Department and Research Center OPTIMAS,
University of Kaiserslautern, 67663 Kaiserslautern, Germany}
\authorinfo{H.C.S.:Email: hcsch@physik.uni-kl.de}

\begin{document}
\maketitle

\begin{abstract}
We present an investigation of electron-spin dynamics in p-doped bulk GaAs
due to the electron-hole exchange interaction, aka the
Bir-Aronov-Pikus mechanism. We discuss under which conditions a spin relaxation times for this mechanism is, in principle, accessible to experimental techniques, in particular to 2-photon photoemission, but also  Faraday/Kerr effect measurements. We give numerical results for the spin relaxation time for a range of p-doping densities and
temperatures. We then go beyond the relaxation time approximation and calculate numerically the
spin-dependent electron dynamics by including the spin-flip
electron-hole exchange scattering and spin-conserving carrier Coulomb
scattering at the level of Boltzmann scattering integrals.  We show that the electronic dynamics deviates from the simple spin-relaxation dynamics for electrons excited at high energies where the thermalization does not take place faster than the spin relaxation time. We also present a derivation of the influence of
screening on the electron-hole exchange scattering and conclude that
it can be neglected for the case of GaAs, but may become important for
narrow-gap semiconductors.

\end{abstract}

\section{Introduction}

Much experimental and theoretical work in the area of spintronics has
been focused on the control and manipulation of the electrons' spin
degree of freedom independently charge. Oftentimes the ultimate goal
is the realization of spintronics devices, in which electronic spins
are the carriers of the
information.~\cite{awschalom:phystoday97,zutic:review} A particularly
appealing property of the manipulation of electronic spins in
semiconductors that sets it apart from existing electronics in
ferromagnetic materials is that the carrier densities can be varied
over several orders of magnitude, and the existence of band gaps makes
a controlled excitation and read-out with optical fields comparatively
easy. The limiting factor for the usefulness of the information
encoded in a spin polarization in a non-ferromagnetic semiconductor is
the spin-polarization decay, which is caused by a variety of
interaction mechanisms~\cite{kikkawa:prl98,heberle:prl01}. A famous
result on the physical limit of spin manipulation by relaxation
phenomena is the measurement of a spin lifetime of about 1 nanosecond
in \emph{n-doped} GaAs at very low
temperatures.~\cite{kikkawa:prl98,heberle:prl01} However, for actual
devices, one might be forced to use materials and excitation
conditions that do not yield these extremely long spin
relaxation-times for several reasons.  First, one would likely prefer
to work with \emph{undoped} semiconductors, or even with dilute
magnetic semiconductors, in which holes are generally the majority
carriers. Second, carrier injection through metal-semiconductor
interfaces or by optical fields may create nonequilibrium electronic
distributions at energies far from the bottom of the
bands~\cite{crooker:apl06-spin-relaxation-times-with-bias}. Third, a knowledge of the spin-dependent
dynamics at moderate to high densities at room temperature is needed
for most devices that interact with optical fields and may be used for
switching of optical signals, such as amplifiers and lasers.

Much work has been done to characterize the spin-flip
interaction mechanisms in semiconductors in terms of spin relaxation
times. State-of-the-art ultrafast optical techniques such as time-resolved Faraday
rotation~\cite{Crooker-prb97} and spin-resolved 2-photon photoemission~\cite{ma:prl97} make it possible to
probe the \emph{spin and energy resolved carrier dynamics} on short
timescales. In this regime, one has to reexamine whether spin relaxation-times constitute a complete description of spin-dependent carrier dynamics, or whether one needs to resort to a microscopic description by including the
relevant scattering processes that change the carrier momentum and
their spin~\cite{wu:prb00:kinetics_qw,glazov-jetplett02,KraussPRL2008:holespin}. 

This paper is devoted to a theoretical study of the
spin-dependent scattering processes due to the
electron-hole exchange interaction in bulk GaAs.  The spin-flip
mechanism due to this interaction, which is the most important one in
semiconductors with high p-doping concentration, is the Bir, Aronov,
and Pikus~(BAP)~\cite{bap:76} mechanism. We first address the question about the conditions under which spin
relaxation-times are appropriate to describe the spin-dependent
dynamics of electrons in p-doped bulk GaAs. We therefore start our presentation in Section~\ref{sec-spin-relaxation-time} with an analysis of how one can define a spin-relaxation time for the BAP process starting from the Boltzmann equation for electron-hole exchange scattering. We then show how this quantity is related to experimental results, and present numerical results for the BAP spin-relaxation time
that are suitable for comparison with experiments over most of the
experimentally interesting density and temperature range.  In Section~\ref{sec-microscopic-dynamics},  we derive the dynamical Boltzmann equations for the electronic distributions including both exchange and direct scattering. This Boltzmann equation was already used in Section~\ref{sec-spin-relaxation-time}, but here we go back to the basic Hamiltonian that describes carriers in semiconductors interacting via the direct Coulomb and the electron-hole exchange interaction. Although these equations have been derived before,
we include a discussion of screening effects in a fashion
analogous with the usual random-phase approximation (RPA) treatment of
screening and we pay special attention to the long-range and short-range contributions to the electron-hole exchange interaction. We then present numerical solutions of the Boltzmann
equation including both the electron-hole exchange interaction and direct electron-hole Coulomb interaction for the case of
the ultrafast optical excitation of a nonequilibrium spin-polarized distribution of electrons and
hole. We compare the the full energy-resolved spin polarization dynamics with the BAP spin relaxation-time.




\section{Spin Relaxation-Time}

\label{sec-spin-relaxation-time}

\subsection{Derivation from Boltzmann equation}

In this Section, we discuss a derivation for the spin relaxation time due to the electron-hole exchange interaction in p-doped semiconductors (GaAs), i.e., the BAP spin relaxation mechanism. We start from the general Boltzmann equation for the electronic distribution functions.
We assume in the following spherically symmetric energy dispersions so that the distribution functions depend only on the carrier energy instead of the vector momentum. We therefore neglect in
this article other spin-relaxing mechanisms, such as the
Dyakonov-Perel mechanism, which depends on anisotropies of the
electronic band structure. We also assume that anisotropic electronic distributions, which are
created by optical excitations are washed out in a few hundred
femtoseconds by carrier-carrier scattering, so that for the timescales
of interest for the electron-hole exchange interaction, we always have
spherically symmetric electronic distributions. With these assumptions we can write the Boltzmann equation for the spin-dependent electronic distributions $n_s(E)$ in the form
\begin{align}
\frac{\partial}{\partial t}n_{s}(E,t)  &  =\int\left\{  -w_{\mathrm{xc}%
}(E,E^{\prime})\left[  1-n_{-s}(E^{\prime})\right]  n_{s}(E)+w_{\mathrm{xc}%
}(E^{\prime},E)n_{-s}(E^{\prime})\left[  1-n_{s}(E)\right]  \right\}
\mathcal{D}(E^{\prime})dE^{\prime}
\label{boltz-E}\\
&  +\int\left\{  -w_{\mathrm{dir}}(E,E^{\prime})\left[  1-n_{s}(E^{\prime
})\right]  n_{s}(E)+w_{\mathrm{dir}}(E^{\prime},E)n_{s}(E^{\prime})\left[
1-n_{s}(E)\right]  \right\}  \mathcal{D}(E^{\prime})dE^{\prime}
\nonumber
\end{align}
Here, $s=\uparrow,$ $\downarrow$ denotes the electron spin projection
quantum number along the quantization axis, which is chosen to be the $z$ axis. The details of the electron-hole exchange and
direct scattering, such as direct carrier Coulomb scattering and/or carrier-phonon scattering, are contained in the scattering kernels which will be discussed in the next section. The exact form of the density of states  
is not important in the following, but for definiteness we consider parabolic electron bands with effective mass $m_{\mathrm{e}}$, so that $\mathcal{D}
(E)=(2m_{\mathrm{e}}/\hbar^{2})^{3/2}/(2\pi)^{2}\times\sqrt{E}$. This equation is derived and solved numerically in the following Section~\ref{sec-microscopic-dynamics}. Instead of a full numerical solution, we analyze first how a
simplified description in terms of a spin-relaxation time is possible,

For the derivation of the spin-relaxation time, we focus on a simple pump-probe scenario.
We assume that such an experiment is performed with a
weak picosecond or sub-picosecond pump pulse that excites electrons close to the bottom of the
band. Then a detailed microscopic description of the coherent carrier-creation
process is not important for the electron spin dynamics because the time scale
for the spin-dependent dynamics is much longer, namely on the order of several hundred picoseconds.
The equilibration of the photoexcited electrons occurs via the direct scattering processes, i.e.,  via Coulomb scattering with equilibrated holes, which are present in the system due to the p-doping, or via electron-phonon scattering. For the present discussion, it is not important which mechanism is dominant, only that the electrons are thermalized after shortly after the pump pulse. If the pump is weak enough, the thermalized electrons will be non-degenerate and described by a Maxwell-Boltzmann distribution
\begin{align}
n_{s}(E,t)  &  =n_{s}(t)f_{\beta}(E)\label{maxboltz}\\
f_{\beta}(E)  &  =N\left(  \frac{2\pi\hbar^{2}}{m_{\mathrm{e}}}\beta\right)^{3/2}
\exp(-\beta E) .
\end{align}
Here, $\beta=1/(k_{B}T)$, $k_{B}$ is Boltzmann's constant, and $N$ is
the total density of electrons.  With this normalization,
$n_{s}(E)$ is in the range of 0 to 1. For this low density of nondegenerate
electrons, the effect
of the direct scattering is to keep the energy dependence of the
electronic distributions constant. This is already built into the
electronic distribution function, Eq.~(\ref{maxboltz}), and its time
development can therefore be calculated by
inserting Eq.~(\ref{maxboltz}) in Eq.~(\ref{boltz-E}), and keeping
\emph{only} the exchange contribution, i.e., the first line, in
Eq.~(\ref{boltz-E}). In addition, for low density distributions as assumed here,
$1-n_{s}(E,t)\approx1$. Integrating the resulting equation over $\int
_{0}^{\infty}$~$\mathcal{D}(E)dE$ and using the normalization
\begin{equation}
\int_{0}^{\infty}dE~\mathcal{D}(E)f_{\beta}(E)=N
\end{equation}
one obtains from Eq.~(\ref{boltz-E}) a dynamical equation for the density of electrons with spin $s$,
\begin{equation}
\frac{\partial}{\partial t}n_{s}(t)= - \langle \gamma_{\mathrm{xc}}
\rangle _{\beta}\bigl[  n_{s}(t)-n_{-s}(t)\bigr]  
\label{dn-average}
\end{equation}
The energy-dependent out-scattering rate $\gamma_{\mathrm{xc}}$, defined by
\begin{equation}
\gamma_{\mathrm{xc}}(E)=\int_{0}^{\infty}w(E,E')\mathcal{D}(E')dE' ,
\label{gamma-xc}
\end{equation}
enters Eq.~(\ref{dn-average}) only in an energy-averaged sense. [See Appendix~\ref{appendix-numerics} for the numerical solution of Eq.~(\ref{gamma-xc})]. The average over the electronic Maxwell-Boltzmann distribution is defined by
\begin{equation}
\left\langle \gamma_{\mathrm{xc}} \right\rangle _{\beta}=\frac{1}{N}\int_{0}^{\infty}
\gamma_{\mathrm{xc}} (E)f_{\beta}(E)\mathcal{D}(E)dE . 
\label{MB-average}
\end{equation}
From Eq.~(\ref{dn-average}) and Eq.~(\ref{maxboltz}), the spin
relaxation time, i.e., the decay time of the spin polarization, defined by
\begin{equation}
P(E,t)=\frac{n_{\uparrow}(E,t)-n_{\downarrow}(E,t)}{n_{\uparrow}
(E,t)+n_{\downarrow}(E,t)},
 \label{spin-pol}
\end{equation}
can be deduced. 
The energy resolved spin
polarization is experimentally accessible, e.g., by spin resolved
2-photon photoemission techniques~\cite{hcs-prb06:spin-relax-surface}. 
However, for low electronic excitation conditions, the energy dependence of the non-degenerate electronic distributions of the form~(\ref{maxboltz}) cancels in Eq.~(\ref{spin-pol}), so that the polarization $P(E)$ is
\emph{energy independent} and equal to the \emph{total} spin polarization
$P(E,t)=P(t)\equiv[ N_{\uparrow}(t)-N_{\downarrow}(t)]/[N_{\uparrow
}(t)+N_{\downarrow}(t)]$. From (\ref{dn-average}) it then follows that
\begin{equation}
\frac{d}{dt}P(t)=-\frac{1}{\tau_{\mathrm{spin}}}P(t) 
\label{dP-dt}
\end{equation}
where the \emph{spin relaxation-time} is defined by
\begin{equation}
\frac{1}{\tau_{\mathrm{spin}}}=2\left\langle \gamma_{\mathrm{xc}}\right\rangle_{\beta}. 
\label{def-taupol}
\end{equation}
For a low density of electrons in an energy range, in which the direct
scattering leads to an effective equilibration on timescales shorter
than the spin relaxation-time, one therefore obtains an \emph{energy
  independent} spin relaxation time. Only this quantity is
experimentally accessible, even with experimental methods that possess an energy resolution that is
better than the spread in energy of the Maxwell-Boltzmann distribution
$f(E)$, which is used to calculate the average in Eq.~(\ref{MB-average}). Moreover, it is impossible to 
measure an energy resolved spin-relaxation time, which is sometimes defined by
\begin{equation}
\frac{1}{\tau_{\mathrm{spin}}(E)}=2\gamma^{\mathrm{xc}}(E)
\label{def-T1-E}
\end{equation}
Thus the only meaningful spin relaxation time is the one originally introduced by Bir, Aronov, and Pikus~\cite{bap:76}, but without the energy dependence, which has been discussed in the literature~\cite{maialle:prb96,maialle:p-doped-holes:prb97}. For the theoretical description of experimental results that do not meet the conditions for the derivation of a single, energy-independent spin relaxation-time, one needs to solve the full Boltzmann equation numerically to obtain the energy or momentum-resolved electron spin dynamics. If the form of the electronic distribution functions changes appreciably over the time of the spin relaxation, one should not expect a single electron spin-relaxation time, but an energy-dependent one. However, from the energy-dependent dynamics one can determine the time evolution accessible by the experiment and try to
fit the spin-polarization dynamics by an exponential decay, if possible. 
Another experimental consequence of this analysis is that a measurement of the electronic spin dynamics by different techniques, such as Faraday rotation, 2-photon photoemission or differential transmission, may yield different results, and one may extract different spin relaxation-times from these measurements, as has been shown for the case of hole spin-relaxation~\cite{KraussPRL2008:holespin}. 

\subsection{Numerical results for spin relaxation-times}

\begin{figure}[b]
\centering
\includegraphics[trim = 3cm 7cm 4cm 8cm, clip, width=0.5\textwidth]{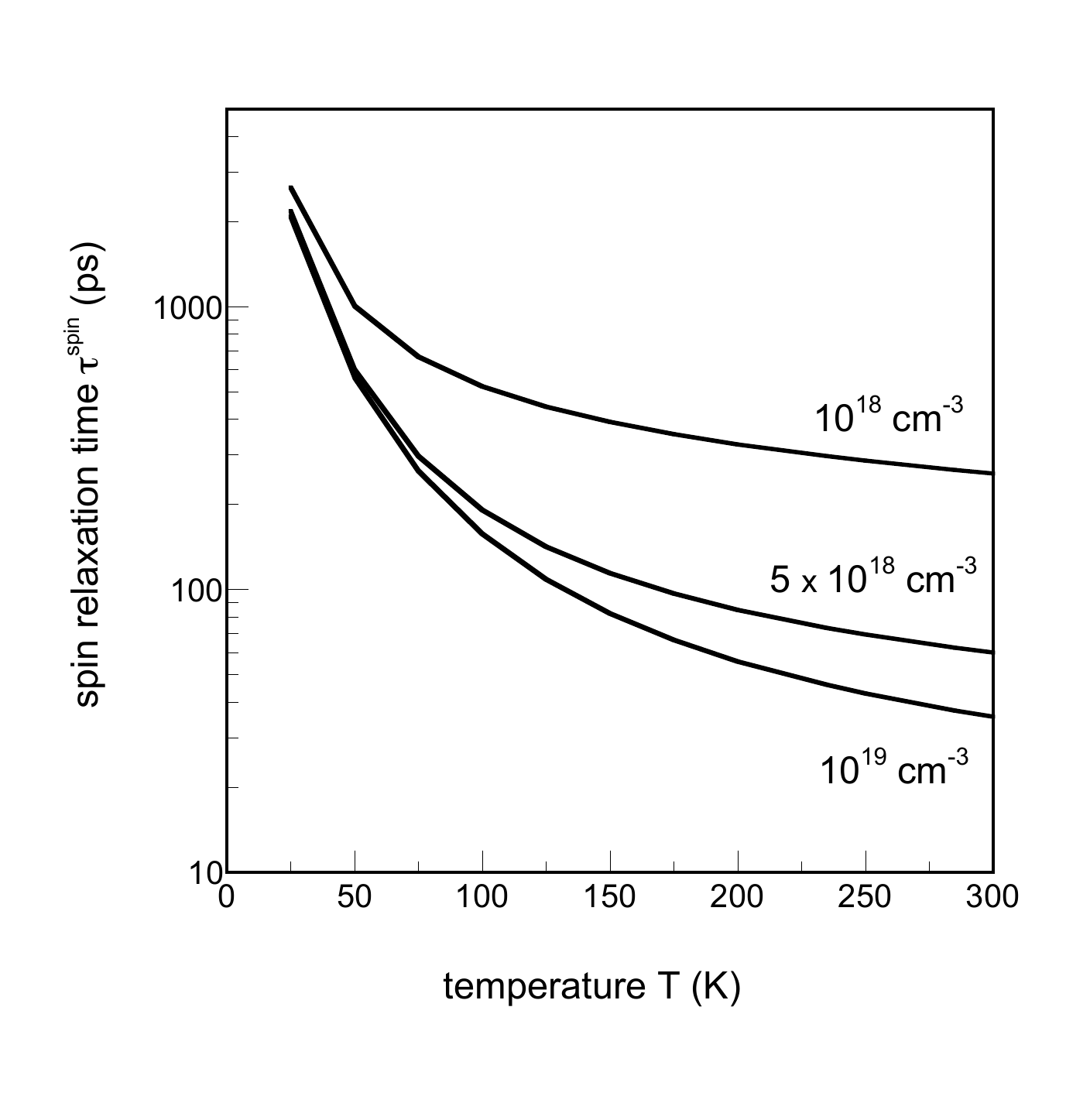}
\caption{Temperature dependence of $\tau_{\mathrm{spin}}$ for different p-doping concentrations.}
\label{tau-vs-T}
\end{figure}

As discussed above, a spin relaxation-time for the BAP process,
Eq.~(\ref{def-taupol}), can be rigorously defined for experiments probing a
low density of thermalized (non-degenerate) electrons. For this case, we numerically evaluate the expression~(\ref{def-taupol}) using also Eqs.~(\ref{MB-average}) and
(\ref{gamma-xc}); see Appendix~\ref{appendix-numerics} and the set of parameters shown in Table
\ref{values} in that Appendix. The results obtained in this Section can therefore be compared to the spin relaxation-times measured by optical pump-probe techniques, such as 2-photon photoemission or
Faraday-rotation measurements, as long as the experiments only probes equilibrated electrons. In this case, as stressed above, the spin relaxation time is energy independent.

An important characteristic of a spin dependent scattering
process is its temperature dependence because different mechanisms differ
strongly in their respective temperature dependences and a measurement of
this quantity can sometimes be used to identify the dominant scattering process for
the material and doping concentration under study. Fig.~\ref{tau-vs-T} shows the
computed spin relaxation-time~$\tau^{\mathrm{spin}}$ from 25\,K up to room temperature. It confirms the well-known general trend that the BAP mechanism becomes important for ``high'' temperatures, i.e., around room temperature, where the spin relaxation-time can reach values below 100\, ps for typical doping densities of $N=5\times 10^{18}$\,cm$^{-3}$ and beyond.

\begin{figure}[tb]
\centering
\includegraphics[trim = 3cm 7cm 4cm 8cm, clip, width=0.5\textwidth]{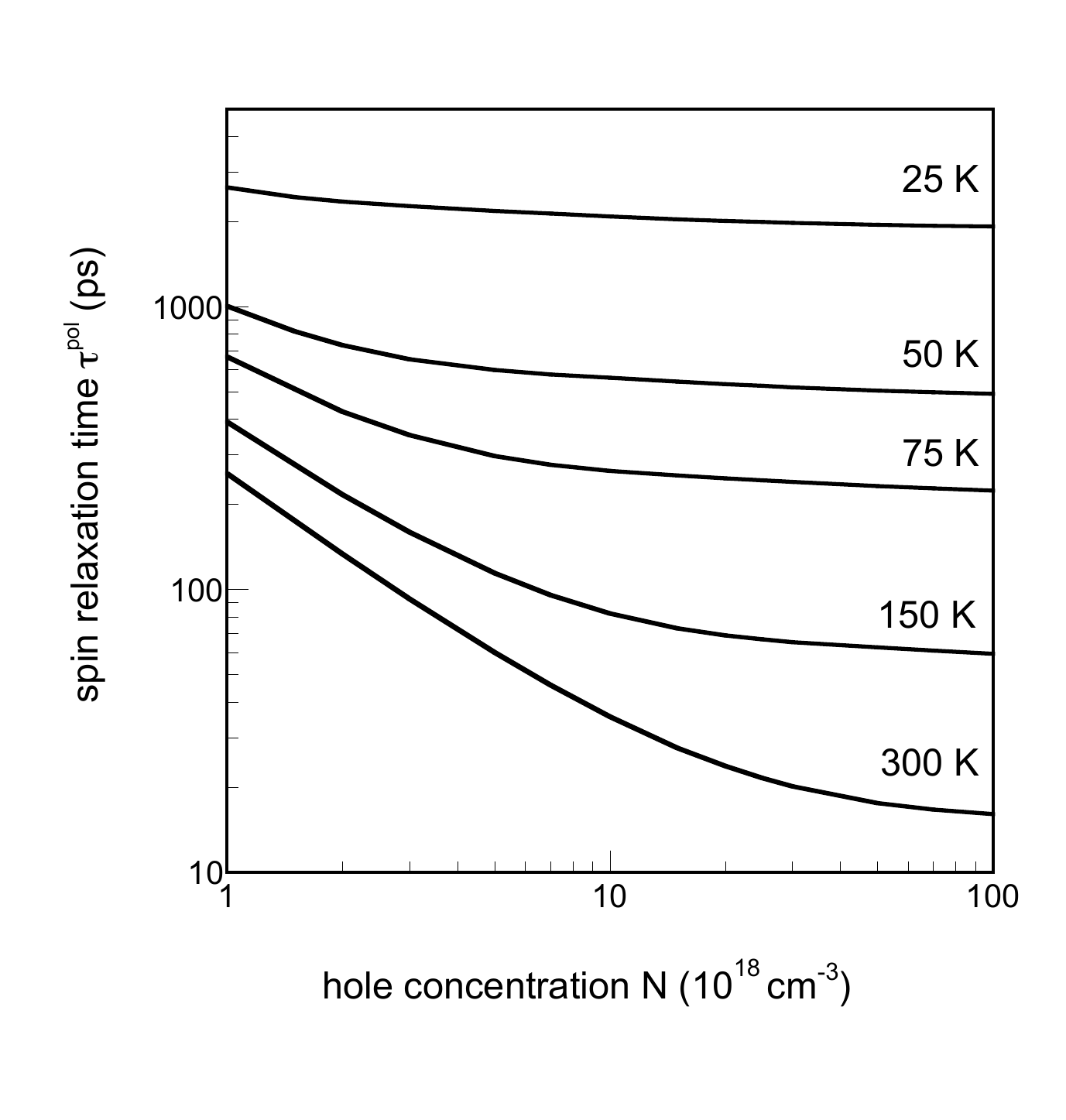}
\caption{Dependence of $\tau_{\mathrm{spin}}$ on the p-doping concentration for different temperatures.}
\label{fig-tau-N}
\end{figure}

Another important ``fingerprint'' of the BAP process is its dependence on the density of holes introduced by the p-doping. The evaluation of Eq.~(\ref{MB-average})  allows one to obtain the BAP spin relaxation-time for arbitrary hole concentrations, as shown in Fig.~(\ref{fig-tau-N}). The doping-density dependence is obtained for the whole density range  without having to patch together analytical results, which can be obtained for the  limiting cases of low and high doping densities. Since both the analytical results are not valid around the densities, for which the holes become degenerate, a discontinuity results if one plots the standard analytical results in one figure over the whole doping density range~\cite{song:prb02}.  One can comare the results of Fig.~\ref{fig-tau-N} with recent evaluations of the contribution of the Dyakonov-Perel effect \cite{JiangWuPRB09:bulk-spin,KraussPRB2010:bulk-GaAs} or with measurements~\cite{oestreich-apl08:bulk-DP} for undoped GaAs, i.e., hole concentrations of well below $10^{17}$\,cm$^{-3}$. For these low densities, even around room  temperature, the BAP process is not the dominant relaxation process. This result shows that there is no ``rule of thumb'' that (for intrinsic GaAs) the Dyakonov-Perel effect is dominant at low temperatures and the BAP becomes important for ``higher'' temperatures. The BAP vs.\ Dyakonov-Perel problem, including the effects of carrier heating, has been explored in more detail recently~\cite{ZhouWuPRB2008,JiangWuPRB09:bulk-spin}.

\begin{figure}[tb]
\centering
\includegraphics[trim = 3cm 7cm 4cm 8cm, clip, width=0.5\textwidth]{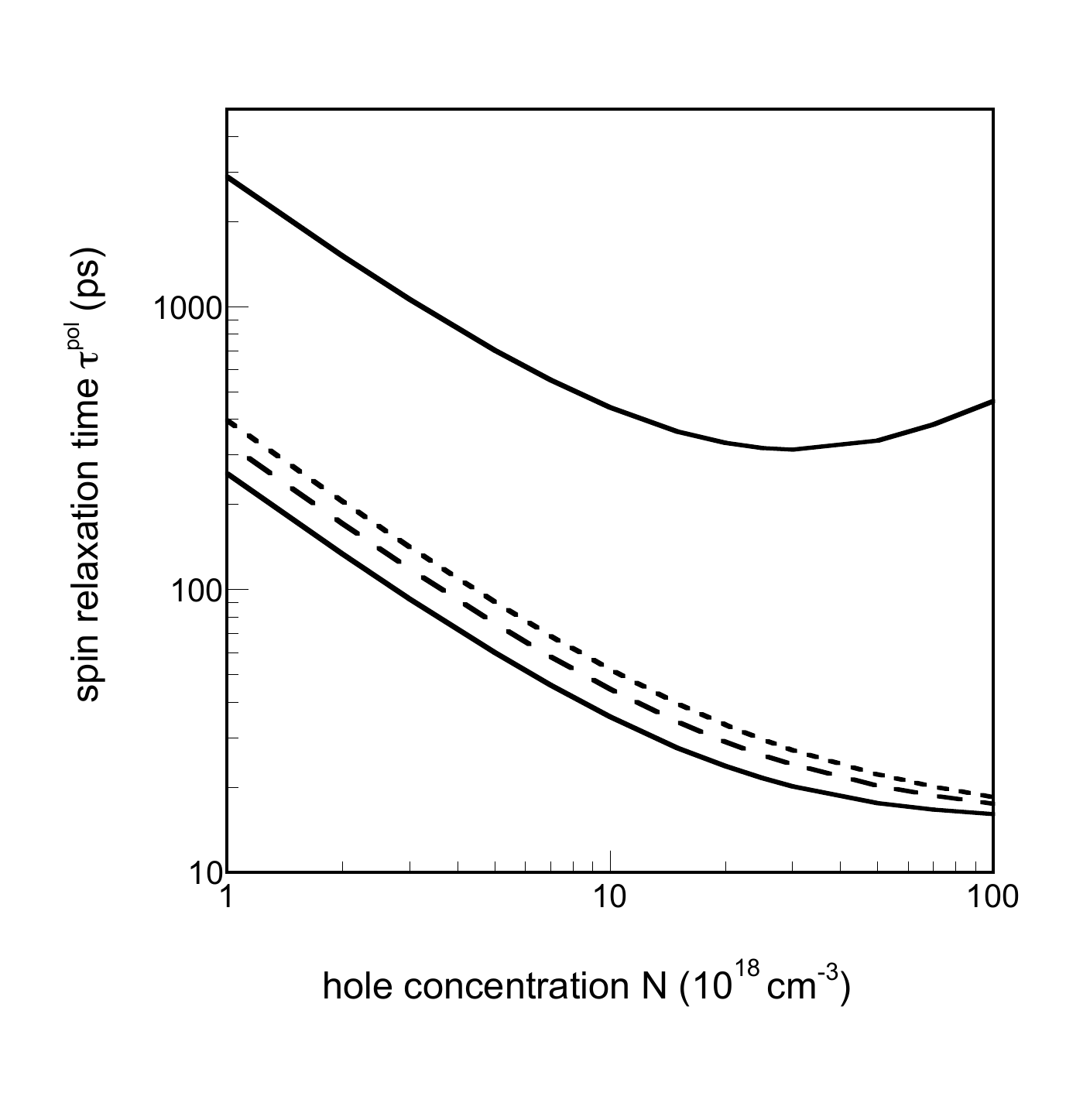}
\caption{Dependence of $\tau_{\mathrm{spin}}$ on the p-doping concentration for $T = 300$\,K. (a) The bottom solid line is the result including short and long range contributions to the electron-hole exchange scattering and contributions from hh-hh, hh-lh, and lh-lh transitions. (b) The long-dashed line is the result including only the long range contribution and all hole bands. (c) The short-dashed is the same as (b), but includes only the hh-hh scattering transitions. (d) The top solid line includes only the short-range contribution to the electron-hole exchange interaction and all relevant transitions between hole bands.}
\label{fig-tau-N-lr}
\end{figure}

In Fig.~\ref{fig-tau-N-lr} we examine the importance of different contributions to the spin-relaxation time due to the electron-hole exchange interaction. The long-range contribution to the electron-hole exchange interaction is clearly the dominant one, as can be seen from comparing the top and bottom solid lines in Fig.~\ref{fig-tau-N-lr}. If one leaves out the short-range contribution one gets a good approximation to the true spin-relaxation time especially for higher p-doping concentrations approaching or exceeding $N = 10^{20}$\,cm$^{-3}$. It is even a good approximation to only include the contribution of the heavy-hole to heavy-hole transitions to the electron spin relaxation-time. Since the results of Fig.~\ref{fig-tau-N-lr} again cover the p-doping density range around hole degeneracy where the approximate results are not valid~\cite{song:prb02}, using just the long-range contribution due to heavy-hole to heavy-hole transitions will give a much better estimate of the true spin relaxation-time than using any one of the approximate expressions.

 
\section{Microscopic Electron Dynamics Due to Exchange Scattering}

\label{sec-microscopic-dynamics}

We now turn towards a derivation of the microscopic spin-dependent carrier dynamics that underlies the concept of the spin relaxation time derived in the previous section. The purpose of this derivation is to make the presentation self-contained and pay attention to the screening properties of the exchange interation. In particular,  all the parameters needed to describe the exchange interaction quantitatively are included in this section and the appendices, and the steps for a numerical solution of the Boltzmann equation are presented in some detail in Appendix~\ref{appendix-numerics}.

\subsection{Hamiltonian and dynamical equations}

Because we have the application to p-doped GaAs in mind, we treat the case of electronic dynamics in the presence of a high density of thermalized holes. We do not include electron-phonon interactions, and do not treat the hole dynamics
and hole-hole Coulomb explicitly. Rather, we assume that the hole
system is always in equilibrium with the lattice, i.e., we have equilibrium hole distributions at the lattice temperature. Furthermore, electron-electron scattering is neglected because electron-hole
scattering is the dominant scattering mechanism due to the low electronic
density. Under these assumptions, the starting point is the basic Hamiltonian
\begin{equation}
H=H_{\mathrm{kin}}+H_{\mathrm{Coul}}+H_{\mathrm{exc}} ,
\label{H}
\end{equation}
which includes the electron kinetic energy and the electron-hole Coulomb
interaction as well as the electron-hole exchange interaction. The kinetic and
the \emph{direct} electron-hole Coulomb term have the familiar
forms~\cite{haug-koch,schaefer-book}
\begin{equation}
H_{\mathrm{kin}}=\sum_{sk}\epsilon_{sk}c_{sk}^{\dag}c_{sk} 
\label{H-kin}
\end{equation}
and
\begin{equation}
H_{\mathrm{Coul}}=\frac{1}{2}\sum_{\vec{k}\vec{k}^{\prime}\vec{q}}
\sum_{sj,s^{\prime}j^{\prime}}2\langle j^{\prime},s^{\prime}|V_{\mathrm{Coul}}
(\vec{q})|j,s\rangle\,c_{j',\vec{k}+\vec{q}}^{\dag}c_{s'\vec{k}'-\vec{q}}^{\dag}c_{s\vec{k}'}c_{j\vec{k}} .
\label{H-coul}
\end{equation}
In Eqs.~(\ref{H-kin}) and (\ref{H-coul}), $c_{s}$, $c_{s}^{\dag}$ and $c_{j}$,
$c_{j}^{\dag}$ are conduction and valence-band destruction and creation
operators, respectively. Conduction-electron states are labeled by the
electron spin projection quantum number $s=\pm1/2$, whereas valence-band
electrons are labeled by the projection quantum numbers $j=\pm1/2$, $\pm3/2$
corresponding to the total angular momentum $J=3/2$. In this paper, we will keep the conduction-valence
(cv) picture for carrier states~\cite{schaefer-book,binder-koch} and the index structure of
direct and exchange terms in the Hamiltonian and throughout the calculations. Only in the final expressions for the scattering kernels a change to
electron and hole distributions is made by replacing $n_{j,k}\longrightarrow
1-n_{j,k}$ and $\epsilon_{j,k}\longrightarrow-\epsilon_{j,k}$. However, the
linear and angular momentum labels, $j$ and $k$, as well as the interaction
matrix elements from the conduction-valence band picture are retained. For a
complete transformation to the electron-hole picture, one would have to change
the angular and linear momenta to $-j$ and $-k$, define hole operators,~\cite{pikus-bir:71} and
calculate the interaction matrix elements using the time-reversed valence-band
wave functions, which leads to a switch of the $j$ and $j'$ labels in
the matrix elements.~\cite{maialle:prb96,maialle:prb93:exciton-spin} The matrix element for the direct Coulomb
interaction is
\begin{equation}
\langle j^{\prime},s^{\prime}|V_{\mathrm{Coul}}(\vec{q})|j,s\rangle
\equiv\delta_{j^{\prime},j}\delta_{s^{\prime},s}v_{q}%
\end{equation}
with the Coulomb potential energy
\begin{equation}
v_{q}=\frac{1}{L^{3}}\frac{e^{2}}{\varepsilon_{0}\varepsilon_{\mathrm{bg}}q^{2}}%
\end{equation}
in SI units, where $\varepsilon_{\mathrm{bg}}$ is the dimensionless background
dielectric constant. The electron-hole exchange interaction is described by
\begin{equation}
H_{\mathrm{exc}}=\sum_{k,k^{\prime},q}\sum_{j,s,j^{\prime},s^{\prime}}\langle
j^{\prime}s^{\prime}|V_{\mathrm{exc}}(\vec{q})|sj\rangle\,
c_{j',\vec{k}+\vec{q}}^{\dag}c_{s^{\prime},\vec{k}^{\prime}-\vec{q}}^{\dag
}c_{j,\vec{k}^{\prime}}c_{s,\vec{k}} . 
\label{H-exc}
\end{equation}
Note in particular that the Hamiltonian of the exchange interaction~(\ref{H-exc})
differs from its direct interaction counterpart~(\ref{H-coul}) in the last two
creation operators and the order of the band indices in the matrix element.
The matrix element of the exchange interaction consists of a
``long range'' and a ``short
range'' part~\cite{pikus-bir:71}
\begin{equation}
\langle j^{\prime},s^{\prime}|V_{\mathrm{exc}}(\vec{q})|s,j\rangle=\langle
j^{\prime},s^{\prime}|V_{\mathrm{exc}}^{\mathrm{lr}}(\vec{q})|s,j\rangle
+\langle j^{\prime},s^{\prime}|V_{\mathrm{exc}}^{\mathrm{sr}}(\vec
{q})|s,j\rangle
\label{def-Vexc}
\end{equation}
where the long range part is given by \cite{pikus-bir:71}
\begin{equation}
\langle j^{\prime},s^{\prime}|V_{\mathrm{exc}}^{\mathrm{lr}}(\vec
{q})|s,j\rangle=v_{q}\frac{\hbar^{2}}{2m_{0}e^{2}}\frac{(\vec{q}\cdot\vec
{p}_{s,j^{\prime}}^{\ast})(\vec{q}\cdot\vec{p}_{s^{\prime},j})}{(\epsilon
_{s}^{0}-\epsilon_{j^{\prime}}^{0})(\epsilon_{s^{\prime}}^{0}-\epsilon_{j}^{0})}
=v_{q}\left(  \vec{q}\cdot\vec{x}_{s,j^{\prime}}^{\ast}\right)  \left(
\vec{q}\cdot\vec{x}_{s^{\prime},j}\right)  
\label{V-lr}
\end{equation}
In Eq.~(\ref{V-lr}), $\vec{p}_{sj}$ is the momentum and $\vec{x}_{sj}$ the dipole operator matrix element between the conduction
electron state $s$, $k=0$ and the valence-band state $j$, $k=0$. The energies $\epsilon^{0}$
are the electronic energies at the $\Gamma$ point, $k=0$, where the conduction
and valence electron bands are respectively degenerate. Using the standard
basis for the coordinate representation of conduction ($S=1/2$) and valence
($J=3/2$) electrons,~\cite{cohen-tannoudji:qm} the short range part can be
brought into the form~\cite{pikus-bir:71} 
\begin{equation}
\langle j^{\prime}s^{\prime}|V_{\mathrm{exc}}^{\mathrm{sr}}(q)|sj\rangle
=\frac{3}{2}\Delta_{\mathrm{exc}}^{\mathrm{sr}}\langle j^{\prime},s^{\prime
}|\left[  \frac{3}{4}+(\vec{J}\cdot\vec{S})\right]  |j,s\rangle\label{V-sr}%
\end{equation}
In Eq.~(\ref{V-sr}), $\Delta_{\mathrm{exc}}^{\mathrm{sr}}$ denotes the
excitonic exchange splitting, and $\vec{J}$ and $\vec{S}$ are the hole and electron angular momentum operators, respectively. Explicit matrix expression for Eqs.~(\ref{V-lr}) and (\ref{V-sr}) are compiled in Appendix~\ref{appendix-matrix-elements}.

In order to describe the scattering dynamics on a microscopic footing, exchange and direct Coulomb interaction need to be included in the equations of motion,
\begin{equation}
\frac{\partial}{\partial t}n_{s}(k)
= \frac{\partial}{\partial t}n_{s}(k)\Bigr|_{\mathrm{Coul}}
+
\frac{\partial}{\partial t}n_{s}(k)\Bigr|_{\mathrm{exc}}
\end{equation}
We therefore present a derivation of
the exchange scattering in a fashion analogous to the
derivation of the Boltzmann equation for direct electron-hole Coulomb scattering in the random-phase approximation
(RPA). Employing a Green's function approach~\cite{schaefer-book,binder-koch}  and using a quasiparticle ansatz
one obtains in the electron-hole picture
\begin{align}
\frac{\partial}{\partial t}n_{s}(k)\Bigr|_{\mathrm{exc}}=& \sum_{\vec{k}^{\prime}}\left[
-W_{k,k^{\prime}}^{\mathrm{exc}}\left(  1-n_{-s,k^{\prime}}\right)
n_{s,k}+W_{k^{\prime},k}^{\mathrm{exc}}n_{-s,k^{\prime}}\left(
1-n_{s,k}\right)  \right]  
\label{Boltz-W} \\
\frac{\partial}{\partial t}n_{s}(k)\Bigr|_{\mathrm{dir}}=& \sum_{\vec{k}^{\prime}}\left[
-W_{k,k^{\prime}}^{\mathrm{dir}}\left(  1-n_{s,k^{\prime}}\right)
n_{s,k}+W_{k^{\prime},k}^{\mathrm{dir}}n_{s,k^{\prime}}\left(
1-n_{s,k}\right)  \right]  
\label{Boltz-W-dir}
\end{align}
where
\begin{align}
W_{k,k^{\prime}}^{\mathrm{exc}}=&\frac{2\pi}{\hbar}\sum_{q}\sum_{j,j^{\prime}%
}|\langle j,-s|\tilde{V}_{\mathrm{exc}}(\vec{q},E_G+\epsilon_{-s,k^{\prime
}}+\epsilon_{j^{\prime},k^{\prime}+q})|sj^{\prime}\rangle|^{2}n_{j,k+q}\left(
1-n_{j^{\prime},k^{\prime}+q}\right) \nonumber\\
& \qquad \qquad \times \delta(\epsilon_{j,k+q}+\epsilon
_{s,k}-\epsilon_{-s,k^{\prime}}-\epsilon_{j^{\prime},k^{\prime}+q})
\label{W-exc} \\
W_{k,k^{\prime}}^{\mathrm{dir}}=&\frac{2\pi}{\hbar}\sum_{q}\sum_{j,j^{\prime}%
}|\langle j,-s|{V}_{\mathrm{scr}}(\vec{q},\epsilon_{s}-\epsilon_{s,k'})|sj^{\prime}\rangle|^{2}n_{j,k+q}\left(
1-n_{j^{\prime},k^{\prime}+q}\right)  \delta(\epsilon_{j,k+q}+\epsilon
_{s,k}-\epsilon_{s,k^{\prime}}-\epsilon_{j^{\prime},k^{\prime}+q})
\label{W-dir}
\end{align}
Note that, instead of writing separate expressions for the in and out-scattering kernels, we have displayed the symmetry of the in and out-scattering processes by introducing only one kernel $W_{k,k'}$. This symmetric form with one kernel is especially convenient for a numerical solution because it avoids the separate calculation of two kernels, which may introduce numerical problems if one violates the balance between in and out scattering, which is responsible for the conservation of the carrier density.

The occurrence of an interband energy in the dynamical exchange interaction is
analogous to the dynamically screened of the direct Coulomb interaction $V_{\mathrm{scr}}$, and can be derived in exactly the same fashion~\cite{schaefer-book,binder-koch}. The result for the
dynamical exchange interaction is
\begin{equation}
\langle js^{\prime}|\tilde{V}_{\mathrm{exc}}(\vec{q},\omega)|sj^{\prime
}\rangle=\langle js^{\prime}|V_{\mathrm{exc}}(\vec{q})|sj^{\prime}\rangle
+\sum_{s_{1}j_{1}}\langle js_{1}|V_{\mathrm{exc}}(\vec{q})|sj_{1}\rangle
\Pi_{s_{1}j_{1}}(\vec{q},\omega)\langle j_{1}s'|\tilde{V}
_{\mathrm{exc}}(\vec{q},\omega)|s_{1}j^{\prime}\rangle
\label{Vs-exc}
\end{equation}
with the retarded RPA electron-hole polarization function
\begin{equation}
\Pi_{sj}(\vec{q},\omega)=-i\hbar\sum_{k}G_{s,k+q}(\omega)G_{j,k}(\omega
)=\sum_{k}\frac{n_{s,k+q}+n_{j,k}}{\hbar\omega-(\epsilon_{s,k+q}%
+\epsilon_{j,k}+E_{G})+i\hbar\gamma} \label{Pi-exc}%
\end{equation}
Here, a zero-density contribution that occurs on transforming into the
electron-hole picture has been subtracted. In Eq.~(\ref{Pi-exc}) an interband
energy enters the denominator. On evaluating Eq.~(\ref{Pi-exc}) for the case of GaAs, 
the product of the exchange interaction and the polarization function in Eq.~(\ref{Vs-exc}) turns out to  so small that the
effect of the screening term on the exchange interaction is
negligible for the numerical results in this paper. 
Therefore only the
unscreened exchange interaction will be used in the following, i.e.,
$\tilde{V}_{\mathrm{exc}}(\vec{q},\omega)\rightarrow V_{\mathrm{exc}}(\vec
{q})$. Even though the effect of screening on the exchange interaction is
comparatively weak for the case of GaAs considered here, it might become more important in narrow band-gap systems. In the following numerical results we will also use a statically screened Coulomb interaction $V_{\mathrm{scr}} (q,\omega) \rightarrow  \varepsilon^{-1}(q)v_q$. A discussion of the static approximation is given by Collet~\cite{collet:93}. Although dynamical screening effects can become important for quantitative results on direct Coulomb scattering, for the purposes of this paper it is mainly important that the direct electron-hole Coulomb scattering occurs much faster than the electron-hole exchange scattering.

The connection to Boltzmann equation~(\ref{boltz-E}) used previously is made in the following way. In the general Boltzmann equation~(\ref{Boltz-W}), which contains distribution functions and energies that may be anisotropic, one assumes spherical symmetry and employs an effective mass
description for electrons and holes. Then
the carrier distributions depend on the energy, i.e.,
\begin{equation}
n_{\nu,k}\rightarrow n_{\nu}(E)\text{ where }E=\frac{\hbar^{2}}{2m_{\ell}}k^{2}
\label{k-to-E}
\end{equation}
for $\nu = $ electrons (e), heavy holes (HH), and light holes (LH).
As mentioned above we assume
further a high density of \emph{equilibrated}, \emph{unpolarized} holes. Then
the distribution functions can be expressed as Fermi functions $n_{j,k}
=f(\epsilon_{k}^{\mathrm{HH}}-\mu)$ for $j=\pm3/2$ and $n_{j,k}=f(\epsilon_{k}^{\mathrm{LH}}-\mu)$ for $j=\pm1/2$ and the $\vec{q}$ summation can be
converted into an integral that can be evaluated mostly analytically as shown
in Appendix~\ref{appendix-numerics}.  The chemical potential $\mu$ is fixed by the temperature and density of the equilibrated holes. Due to the assumption of unpolarized holes and the symmetry
properties of the interaction matrix element, which are discussed in the
Appendix~\ref{appendix-matrix-elements}, the kernel~\ref{W-exc} is a function $W(E,E')$ independent of the electron spin. Writing the wave-vector summations in Eq.~(\ref{Boltz-W}) as energy integrals including the density of states, one finally obtains Eq.~(\ref{boltz-E}), on which the analysis of the BAP spin-relaxation time is based.


\subsection{Numerical Results for Spin-Polarization Dynamics}

In this Section, the dynamics of spin polarized electrons under the influence
of exchange (spin-flip) and direct (non spin-flip) electron-hole scattering processes are
numerically evaluated using Eqs.~(\ref{Boltz-W}), (\ref{W-exc}) and the static (non-screened) exchange interaction matrix element~(\ref{def-Vexc}). Here, we analyze the energy and spin-resolved electron dynamics for excitation conditions that would be used in experiments designed to measure the
energy-resolved polarization dynamics. One of our goals here is to show, which
energy-resolved quantities can be extracted from typical experiments. For
experiments with sub-picosecond excitation pulses, the a detailed microscopic
description of the coherent carrier-creation process is again neglected because it occurs on a shorter time scale than  for the the spin-dependent dynamics. As a model for such an experiment that
creates a nonequilibrium electron distribution by a short laser pulse, we
consider as an initial condition a Lorentzian electron distribution peaked
around an electronic energy $\epsilon_{k_{L}}$, or modulus of the electron
wave vector $k_{L}$. In an experiment, the wave vector $k_L$ and the energy would be determined by the laser wavelength and the band structure~\cite{hilton-tang,hcs-prb06:spin-relax-surface}. We restrict ourselves to excited electron densities that are small enough so that the neglect of the electron-electron interactions is justified. The initial low-density (nondegenerate) electron distribution is chosen with initial spin polarization of 0.5 and with the same
temperature as the holes.

\begin{figure}[b]
\centering
\includegraphics[trim = 3cm 10cm 4cm 9cm, clip, width=0.45\textwidth]{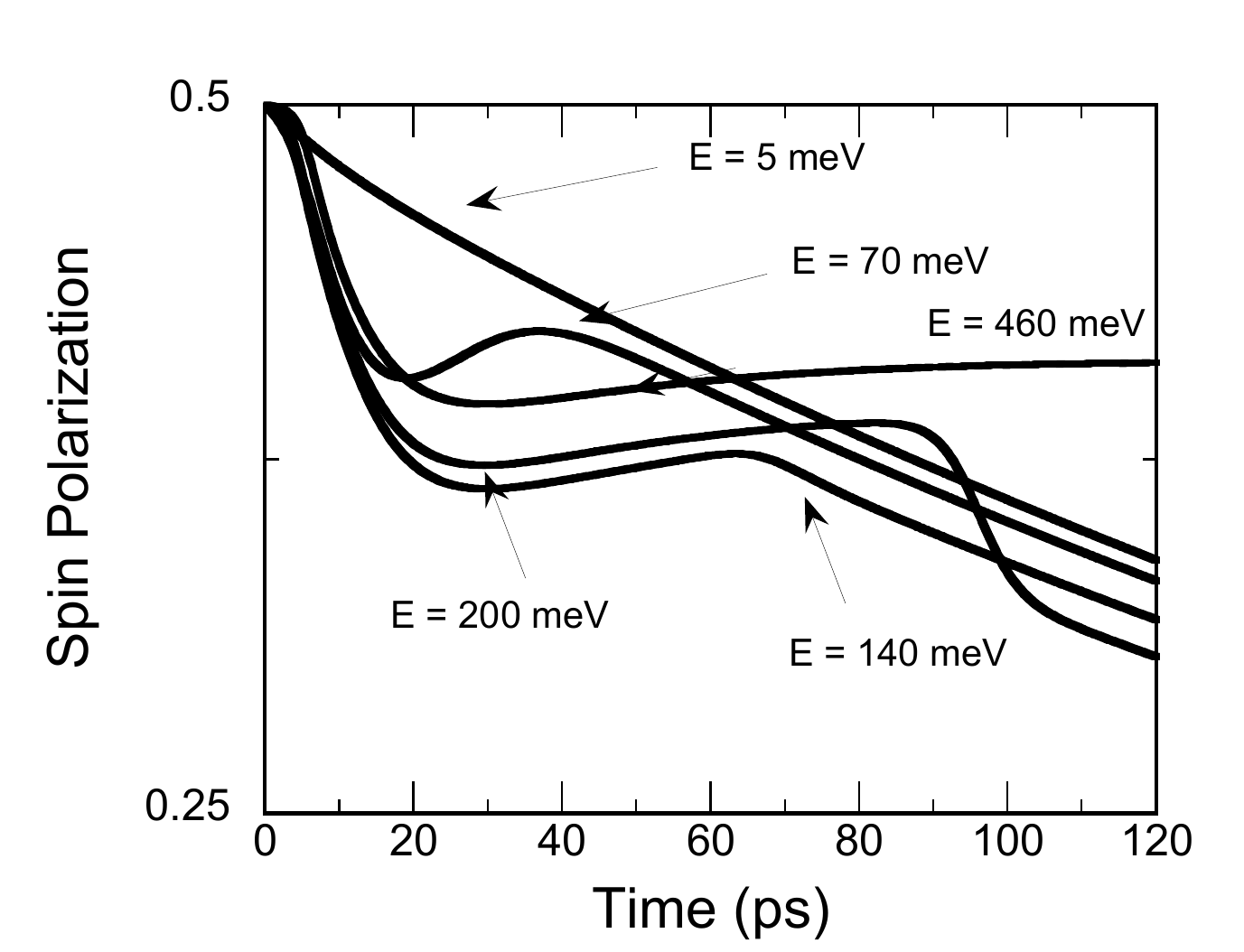}
\caption{Spin polarization dynamics for a p-doping concentration of $5\times 10^{18}$\,cm$^{-3}$ and temperature $T= 75$\,K for excitation of electrons at different energies $E$ above the bottom of the band.}
\label{fig-poldynamics-75K}
\end{figure}

Figure~\ref{fig-poldynamics-75K} shows the dynamical results for the spin polarization $P(E,t)$ defined
in Eq.~(\ref{spin-pol}). The curves were calculated numerically as
described in Appendix~\ref{appendix-numerics} with both the short-range and
the long-range contribution to the electron-hole exchange interaction matrix
element taken into account. The parameters are again those given in Table~\ref{values}. We assume unpolarized holes with density
$N_{\mathrm{A}}=5\times10^{18}\,\mathrm{cm}^{-3}$ and temperature
$T_{\mathrm{h}}=75$\,K. The spin polarization is monitored at the energy $E$, at which the carriers are initially created in the form of a nonequilibrium electron distribution peaked at $E$.   At a carrier energy
$E=5$\,meV, i.e., close to the bottom of the electron band, the decay of the spin polarization after the excitation of the electrons is
almost exponential. For higher peak energies of the excited electrons, already starting below an excess energy of 70\,meV, the dynamics is not described by an exponential drop at all. Instead, the $P(E,t)$ rises after an initial steep drop.
Only for longer times, the spin polarization dynamics resembles
an exponential decay with time constants of about 305\,ps, which is identical to the time constant of the decay of the 
total spin polarization and almost identical to the BAP spin relaxation-time as shown in Figs.~\ref{tau-vs-T} and~\ref{fig-tau-N}. The reason for this behavior is the interplay between
the spin-flip exchange scattering and the direct scattering. If the polarization is determined at an energy where electrons undergo in and out-scattering processes, a quasi-equilibrium is reached due to direct electron-hole scattering events on a timescale faster than the spin-flip
scattering, so that the spin-flip scattering occurs for effectively equilibrated
electrons. For these, the concept of a spin relaxation time exists, as shown in Section~\ref{sec-spin-relaxation-time}. For electrons with higher energies out-scattering
processes, dominate for several ten picoseconds
up to more than 100\,ps. On this timescale, the initial peaked electronic distribution has been already been washed out, but no quasi-equilibrium has been established
for electrons at these energies. The polarization dynamics is therefore due to a combination of spin-conserving and spin-flip scattering processes, and the dynamics depend on the form of the actual nonequilibrium electron
distribution. This non-exponential dynamics occurs until a quasi-equilibrium is established due to direct scattering
processes. Then the spin polarization decay becomes exponential.  Thus experiments exciting and probing the electronic spin polarization at higher energies will not yield energy-resolved information about the the energy resolved spin-relaxation time due to the exchange interaction as defined in Eq.~(\ref{def-T1-E}).

\begin{figure}[t]
\centering
\includegraphics[trim = 3cm 10cm 4cm 9cm, clip, width=0.45\textwidth]{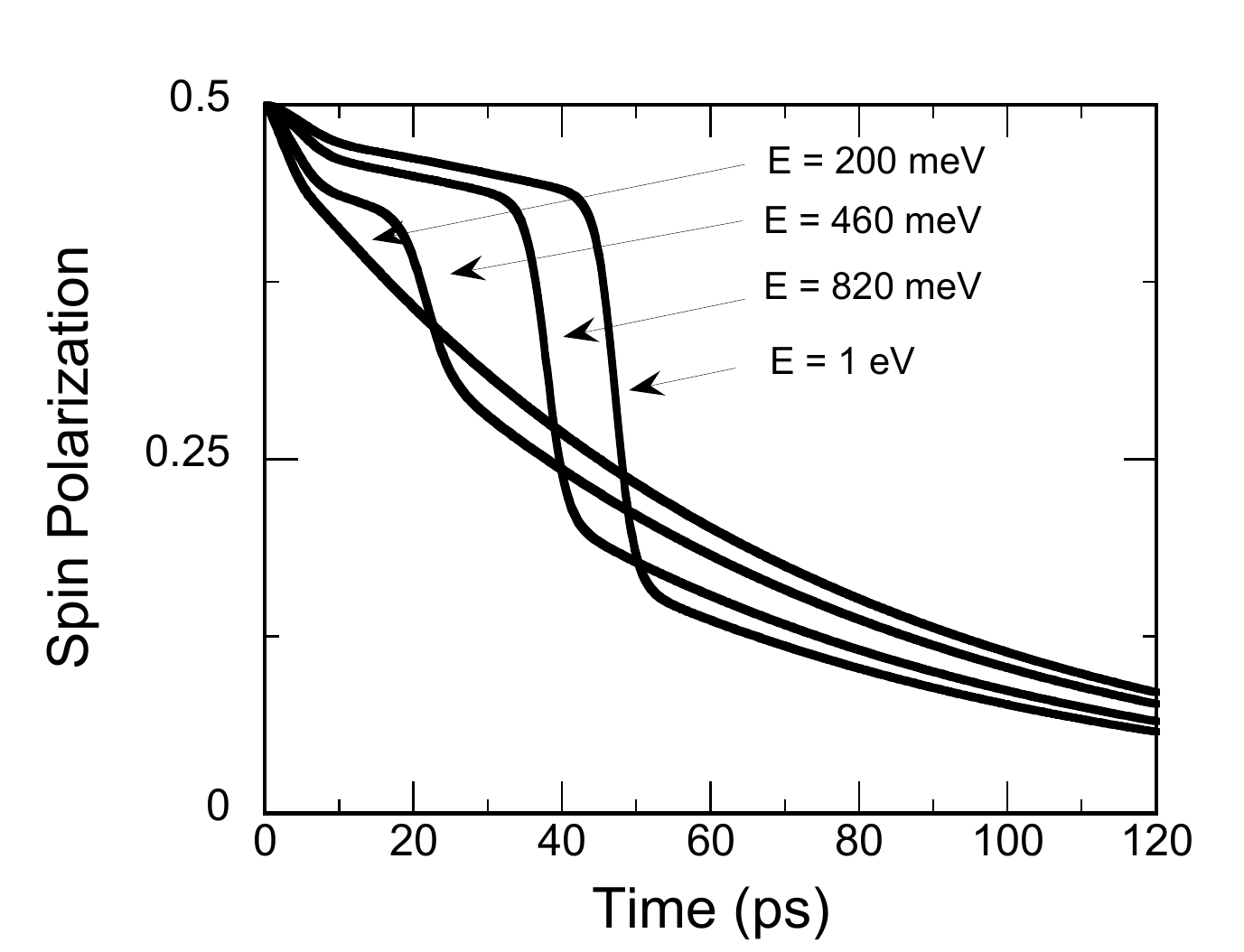}
\caption{Spin polarization dynamics for a p-doping concentration of $5\times 10^{18}$\,cm$^{-3}$ and temperature $T= 300$\,K for excitation of electrons at different energies $E$ above the bottom of the band.}
\label{fig-poldynamics-300K}
\end{figure}

Figure~\ref{fig-poldynamics-300K} shows the same scenario for a hole temperature of $T= 300$\, K. In this case, again, the spin polarization decays exponentially only for electrons close to the bottom of the band, with a spin-relaxation time of about 70\,ps, which is again almost identical to the spin-relaxation time of the total polarization and the BAP spin relaxation time. At higher energies, the initial polarization dynamics is non-exponential. However, the onset of the nonexponential dynamics occurs only for electrons with much higher excess energy, here for energies well above 200\,meV because at higher energies the equilibrated ``tail'' of the distributions stretches out to higher energies. This indicates that the nonexponential dynamics is not as important and will not be as easily observed as it is the case for $T=75$\,K. The dynamics generally occur on a shorter time scale than in the $T=75$\, K case (note the different $y$ axes in the two figures), but the ``deviation'' from the exponential decay is qualitatively different. Now the nonequilibrium dynamics leads to a \emph{slower} initial decay of the polarization, before the dynamics becomes exponential when a quasi-equilibrium is established by the direct scattering processes. Since the initial dynamics is so much different for the different temperatures there is no simple rule of thumb which qualitative behavior of the spin dynamics occurs for nonequilibrium conditions. Certainly, the spin relaxation-time cannot serve even as guideline in this case, and one has to model the excitation conditions and energy-dependent measurement accurately to obtain reliable results that may be compared to a given experiment.

\section{Conclusion}
We have investigated how spin relaxation-times can be used to describe the spin polarization dynamics of electrons due to the electron-hole exchange interaction, i.e., the Bir-Aronov-Pikus mechanism. It is found that there is only one \emph{ensemble averaged} spin relaxation time that can be accessed by optical experiments, such as Faraday rotation, differential transmission, or 2-photon photoemission, even if these experiments have an energy resolution that is smaller than the spread of excited electronic energies. In order to derive a meaningful spin-relaxation time, one has to assume that a nondegenerate electronic distribution is kept a quasi-equilibrium by the direct scattering processes while the exchange scattering effects the spin polarization. Thus the energy-dependence of the spin relaxation dynamics is washed out by the thermal average. If one tries to get around measuring the ensemble average and intends to probe the electronic spin polarization at energies, at which there is initially no quasi-equilibrium established, one still does not get a unique energy dependent spin relaxation-time, but rather a complicated spin polarization dynamics that is determined by the interplay of direct and spin-flip scattering processes. It is therefore not possible to extract an energy-dependent spin relaxation-time because the of the direct scattering mechanisms, although an energy-dependent energy relaxation time can be defined, as is usually done for electronic lifetimes.


\begin{acknowledgments}
We thank M. Aeschlimann and  W. H\"ubner for helpful discussions and acknowledge financial support by the DFG through the Graduiertenkolleg 792 ``Nonlinear Optics and Physics on Ultrashort Time Scales.''
\end{acknowledgments}

\appendix

\section{Matrix Elements}
\label{appendix-matrix-elements}

The exchange interaction between electrons and holes consists of a short-range
and long-range contribution In the following, electron-hole states are labeled
by conduction-valence band angular momentum labels $(js)=$ ($-\frac{3}
{2}\uparrow),(-\frac{3}{2}\downarrow),(-\frac{1}{2}\uparrow),(-\frac{1}
{2}\downarrow),(+\frac{1}{2}\uparrow),(+\frac{1}{2}\downarrow),(+\frac{3}
{2}\uparrow),(+\frac{3}{2}\downarrow)$

The \emph{long-range part} to the exchange interaction matrix results from the
evaluation of Eq.~(\ref{V-lr}) in the standard angular-momentum basis and
reads:
\begin{equation}
V_{\mathrm{exc}}^{\mathrm{lr}}(\vec{q})=v_{q}|r_{cv}|^{2}\left(
\begin{array}
[c]{cccccccc}%
0 & 0 & 0 & \frac{1}{2\sqrt{3}}q_{\perp}^{2} & 0 & \frac{1}{\sqrt{3}}
q_{+}q_{z} & 0 & -\frac{1}{2}q_{+}^{2}\\
0 & \frac{1}{2}q_{\perp}^{2} & 0 & \frac{1}{\sqrt{3}}q_{z}q_{+} & 0 &
-\frac{1}{2\sqrt{3}}q_{+}^{2} & 0 & 0\\
0 & 0 & \frac{1}{6}q_{\perp}^{2} & \frac{1}{3}q_{z}q_{-} & \frac{1}{3}
q_{+}q_{z} & \frac{2}{3}q_{z}^{2} & -\frac{1}{2\sqrt{3}}q_{+}^{2} & -\frac
{1}{\sqrt{3}}q_{z}q_{+}\\
\frac{1}{2\sqrt{3}}q_{\perp}^{2} & \frac{1}{\sqrt{3}}q_{z}q_{+} & \frac{1}
{3}q_{z}q_{+} & \frac{2}{3}q_{z}^{2} & -\frac{1}{6}q_{+}^{2} & -\frac{q_{z}
}{3}q_{+} & 0 & 0\\
0 & 0 & \frac{1}{3}q_{-}q_{z} & -\frac{1}{6}q_{-}^{2} & \frac{2}{3}q_{z}^{2} &
-\frac{1}{3}q_{-}q_{z} & -\frac{1}{\sqrt{3}}q_{+}^{2}q_{z} & \frac{1}%
{2\sqrt{3}}q_{\perp}^{2}\\
\frac{1}{\sqrt{3}}q_{-}q_{z} & -\frac{1}{2\sqrt{3}}q_{-}^{2} & \frac{2}%
{3}q_{z}^{2} & -\frac{1}{3}q_{-}q_{z} & -\frac{1}{3}q_{+}q_{z} & \frac{1}%
{6}q_{\perp}^{2} & 0 & 0\\
0 & 0 & -\frac{1}{2\sqrt{3}}q_{-}^{2} & 0 & -\frac{1}{\sqrt{3}}q_{-}q_{z} &
0 & \frac{1}{2}q_{\perp}^{2} & 0\\
-\frac{1}{2}q_{-}^{2} & 0 & -\frac{1}{\sqrt{3}}q_{-}q_{z} & 0 & \frac
{1}{2\sqrt{3}}q_{\perp}^{2} & 0 & 0 & 0
\end{array}
\right)
\label{Vlr-matrix}
\end{equation}
Here, $q_{\perp}^{2}=q_{x}^{2}+$ $q_{y}^{2}$ is modulus of the transverse
component of $\vec{q},$ and $q_{z}$ is the $z$ component. Then
\begin{equation}
q_{\pm}=q_{x}\pm\mathrm{i}q_{y}=q_{\perp}e^{\pm\mathrm{i}\varphi}%
\end{equation}
where $\varphi$ is the azimuthal angle of $\vec{q}.$

Pikus and Bir~\cite{pikus-bir:71} also show how the \emph{short-range contribution} arises directly from the Coulomb interaction. Here, one evaluates the following exchange Coulomb matrix elements
\begin{equation}
\langle j'\vec{k}_{1},s'\vec{k}_{2}|v|s\vec
{k}_{3},j\vec{k}_{4}\rangle=\sum_{\sigma,\sigma^{\prime}}\int
\varphi_{j'\vec{k}_{1}}(\vec{x}\sigma)^{\ast}\varphi
_{s'\vec{k}_{2}}(\vec{x}^{\prime}\sigma^{\prime})^{\ast}v(\vec{x}-\vec{x}')\varphi_{s\vec{k}_{3}}(\vec{x}%
\sigma)\varphi_{j\vec{k}_{4}}(\vec{x}^{\prime}\sigma^{\prime})d^{3}%
xd^{3}x'
\end{equation}
The carrier states are of the usual form
\begin{equation}
\varphi_{\alpha\vec{k}}(\vec{x},\sigma)=\frac{1}{\sqrt{L^3}}\exp(-i\vec{k}\cdot\vec{x})u_{\alpha}(\vec{x}\sigma),
\end{equation}
where $u_{\alpha}(\vec{x}\sigma)$ are the Bloch functions at $k=0$. For electrons, $\alpha = s$, and for holes, $\alpha = j$, so that the angular momentum dependence of the $u$ is determined by their angular momentum quantum number.

The contribution due of the exponential factors can be treated as in the spin-diagonal Coulomb matrix element by dividing the integral over the crystal into integrals over unit cells and pulling the exponential factors out of the unit cell integrals. The remaining integrals over the unit cell are evaluated by inserting the explicit expressions for the $u$s, whose angular momentum dependence is given by their total angular momentum projection quantum number $j$, i.e., by linear combinations of spherical harmonics. Expanding the interaction $v(\vec x - \vec x') = \sum_\ell V_\ell(r,r') \sum_{m=-\ell}^{\ell} Y^\ell_m(\hat{x}_1)Y^{\ell}_m(\hat{x}_2)^*$, the angular momentum integrals can be calculated and the remaining integrals over $r$ and $r'$ can be absorbed into a quantity that is accessible experimentally. One finds in the same matrix representation as Eq.~(\ref{Vlr-matrix})
\begin{equation}
V_{\mathrm{exc}}^{\mathrm{sr}}=\frac{1}{L^{3}}\left(
\begin{array}
[c]{cccccccc}%
0 & 0 & 0 & \frac{1}{2}\sqrt{3} & 0 & 0 & 0 & 0\\
0 & \frac{3}{2} & 0 & 0 & 0 & 0 & 0 & 0\\
0 & 0 & \frac{1}{2} & 0 & 0 & 1 & 0 & 0\\
\frac{1}{2}\sqrt{3} & 0 & 0 & 1 & 0 & 0 & 0 & 0\\
0 & 0 & 0 & 0 & 1 & 0 & 0 & \frac{\sqrt{3}}{2}\\
0 & 0 & 1 & 0 & 0 & \frac{1}{2} & 0 & 0\\
0 & 0 & 0 & 0 & 0 & 0 & \frac{3}{2} & 0\\
0 & 0 & 0 & 0 & \frac{1}{2}\sqrt{3} & 0 & 0 & 0
\end{array}
\right)  \cdot\frac{3}{2}V_{\text{uc}}\Delta_{\text{SR}}
\label{Vsr-matrix}
\end{equation}
where
\begin{equation}
\Delta_{\text{SR}}=\sum_{\sigma,\sigma^{\prime}}\int_{\mathrm{unit~cell}%
}u_{\alpha^{\prime}}(\vec{x}\sigma)^{\ast}u_{\beta^{\prime}}(\vec{x}^{\prime
}\sigma^{\prime})^{\ast}\left[v(\vec{x}-\vec{x}')\right]
_{\mathrm{periodic}}u_{\alpha}(\vec{x}\sigma)u_{\beta}(\vec
{x}^{\prime}\sigma^{\prime})d^{3}xd^{3}x^{\prime} \label{Delta-sr}%
\end{equation}
is the exchange splitting energy, $V_{\text{uc}}$ is the volume of the unit
cell and $L^{3}$ the crystal volume. The Coulomb potential $v$ appearing in Eq.~(\ref{Delta-sr}) is lattice periodic because it results from a summation over reciprocal lattice vectors. The energy $\Delta_{\text{SR}}$ is
related to the 1s-exciton splitting (into one triplet and one quintuplet)
$\Delta_{\text{SR}}^{\text{exciton}}$ due to the exchange interaction by
\begin{equation}
V_{\text{uc}}\Delta_{\text{SR}}=\frac{1}{2}\pi a_{B}^{3}\Delta
_{\text{\textrm{SR}}}^{\text{\textrm{exciton}}}, 
\label{sr-exciton}
\end{equation}
where $a_{B}$ is the excitonic Bohr radius. 

Using the matrix expressions (\ref{Vlr-matrix}) and (\ref{Vsr-matrix}) in (\ref{Boltz-W}) one calculates the squared matrix elements for electronic spin-flip scattering due to transitions between the different hole bands. For instance, for heavy-hole to heavy-hole scattering one has
\begin{equation}
|\langle 3/2,\downarrow|V_{\mathrm{xc}}(\vec
{q})|\uparrow,3/2 \rangle|^{2} = \left\vert v_q|r_{cv}|^{2}(-\frac{1}{2}q_{+}^{2})\right\vert ^{2}
= \left(
\frac{e^2}{\varepsilon_{0}\varepsilon q^2} |r_{cv}|^{2} q_\perp^{2}\right)^{2}  
=\alpha^{2}\left(
1-z^{2}\right)  ^{2}
\label{hh}
\end{equation}
where $\alpha=\frac{e^{2}}{\epsilon_{0}\epsilon}|r_{\mathrm{cv}}|^{2}$, and we have defined $z = q_z/q$ so that $q_\perp^2 = q^2 - q_z^2 \equiv q^2(1-z^2)$.
In a similar fashion one obtains for heavy-hole to light-hole scattering (two contributions in the scattering kernel)
\begin{equation}
\frac{1}{3}\left(  \frac{\alpha}{2}(1-z^{2})+\frac{3}{2}\Delta_{\text{SR}}\right)  ^{2}
\label{lh}
\end{equation}
and for light-hole to light-hole scattering
\begin{equation}
\left(\frac{2}{3}z^{2}\alpha+\Delta_{\text{SR}}\right)^{2}+(1-z^{2})\frac{\alpha^{2}}{36} 
\label{ll}
\end{equation}
The full squared interaction matrix element in Eq.~(\ref{Boltz-W}) is a sum of Eqs.~(\ref{hh})--(\ref{ll}), but the heavy-hole to heavy-hole contribution~(\ref{hh}) is the dominant one for the conditions investigated in this paper. Note that there is no dependence of the in-plane angle $\varphi$ (between $q_{x}$ and
$q_{y}$) for matrix elements that also have a short-range contribution.
Therefore there is no explicit $\varphi$ dependence in the terms that have
combined short-range and long-range contributions.

\begin{table}[t] \centering
\begin{tabular}
[c]{|l||l|l|}\hline
Physical quantity & Symbol & Value\\\hline\hline
Dipole matrix element & $d_{\text{cv}} = e\,r_{\text{cv}}$ & $0.6e\,\mathrm{nm}$\\\hline
Excitonic exchange splitting & $\Delta_{\text{SR}}^{\text{exciton}}$ &
$0.05\,\mathrm{meV}$\\\hline
Electron effective mass & $m_{\mathrm{e}}$ & 0.067\\\hline
Heavy hole effective mass & $m_{\mathrm{HH}}$ & 0.457\\\hline
Light hole effective mass & $m_{\mathrm{LH}}$ & 0.087\\\hline
Background dielectric constant & $\varepsilon_{bg}$ & 12.5\\\hline
\end{tabular}
\caption{Values of constants used in the numerical calculations.\label{values}}
\end{table}

\section{Numerical evaluation of BAP spin relaxation-time}

\label{appendix-numerics}

To obtain the BAP spin relaxation-time one evaluates the expression, with momentum indices chosen for numerical convenience and as a function of $k$ instead of $E$, but cf.~Eq.(\ref{k-to-E}),
\begin{equation}
\gamma_{\mathrm{xc}}(k) =\frac{2\pi}{\hbar}\sum_{\vec{q},\vec
{p}}\sum_{j,j^{\prime}}|\langle j^{\prime},-s|V_{\mathrm{xc}}(\vec{q})|s,j\rangle|^{2}\left(  1-n_{j,\vec{p}+\vec{q}}\right)  n_{j',\vec{k}+\vec{q}}\
\delta(\epsilon_{k}^{e}+\epsilon_{\vec{k}+\vec{q}}^{j}
-\epsilon_{\vec{p}}^{e}-\epsilon_{\vec{p} + \vec{q}}^{j'}). 
\label{spindecay}
\end{equation}
First, the momentum sums over $\vec{q}$ and $\vec{p}$ are
replaced by two three-dimensional integrals using spherical coordinates
with the $p$ integration as the inner integral. One measures
$p$ against $q,$ and $q$ against $k,$ denoting $z'
=\cos\angle(\vec{p},\vec{q})$ and $z=\cos$ $\angle(\vec{q},\vec{k})$ together
with their respective azimuthal angles $\varphi'$ and $\varphi$. The
$\varphi'$ and $\varphi$ integrations yield only $2\pi$ each because
there is no explicit dependence on these angles in the momentum labels
$\vec{k}+\vec{q}$ or $\vec{k}'-\vec{q}$, and the exchange matrix
element does not depend on $\varphi'$, only on $q$ and $z'$.
Then one analytically evaluates the integral over $z'$ making use of the
properties of $\delta$ distribution. For
these manipulations, it is convenient to introduce a unit of
length, denoted by $\ell$, and make the integrations over $q$ and $k^{\prime}$
dimensionless by defining $\tilde{q}=q\ell$ and $\tilde{k}^{\prime}=$
$\tilde{k}\ell$. It is also convenient to introduce the energy
\begin{equation}
E_{\mathrm{kin}}=\frac{\hbar^{2}}{2m_{0}\ell^{2}}
\end{equation}
to make the kinetic energies dimensionless, $\epsilon^{j,e}_p/ E_{\mathrm{kin}}= \tilde{p}^2/m_{j,e}$. The masses are expressed in terms of the vacuum electron mass $m_0$. The result is (expressed  in our dimensionless units and leaving the tildes off)
\begin{equation}
\gamma_{\mathrm{xc}}(k)
=\frac{1}{32\pi^{3}}\frac{1}{\hbar\beta E_{\mathrm{kin}}^2}
\int_{0}^{\infty}dqq\int_{-1}^{1}dz \,
\sum_{jj'} f\left(\epsilon^j_{\sqrt{k^2 + q^2 + 2kqz}}-\mu\right) 
|\langle j',\downarrow|V_{\mathrm{xc}}(z)|\uparrow,j \rangle|^{2}
m_e m_{j'} \log\left( \frac{1+ e^{\zeta_{\mathrm{max}}}}{1-e^{\zeta_{\mathrm{min}}}}\right) 
\end{equation}
In this equation, the squared interaction matrix elements~Eqs.~(\ref{hh})--(\ref{ll}) are used, and we have used the following definitions.
First, the Fermi-Dirac funtion $f(\xi) = [1+\exp(\beta\xi)]$ and the hole chemical potential~$\mu$, further
\begin{equation}
\zeta_{\mathrm{min,max}} = \beta \left( \frac{1}{m_{j'}} p^2_{\mathrm{min,max}} - \mu_{j'} \right)
\end{equation}
where $\mu_{j'}$ is the reduced mass of an electron and a hole in band $j'$. Further,
\begin{equation}
p_{\mathrm{min,max}}  =\left\vert \sqrt{\mu\left[\frac{1}m_{j} {(k^2+q^2+2kqz)}+ \frac{1}{m_{\mathrm{e}}}k^2 
- \frac{M_{j'}}{m_{\mathrm{e}}}q^2\right]}-\frac{\mu_{j'}}{m_{\mathrm{e}}}q\right\vert 
\end{equation}
with $M_{j'} = m_{\mathrm{e}} + m_{j'}$.

\bibliographystyle{spiebib}
\bibliography{dephasing2,spin3,hcs_pubs2}

\begin{thebibliography}{10}

\bibitem{awschalom:phystoday97}
Kikkawa, J.~M. and Awschalom, D.~D., ``Electron spin and optical coherence in
  semiconductors,'' {\em Physics Today}~{\bf 32},  33 (1999).

\bibitem{zutic:review}
{\v{Z}}uti{\'c}, I., Fabian, J., and Das~Sarma, S., ``Spintronics: Fundamentals
  and applications,'' {\em Reviews of Modern Physics}~{\bf 76},  323
  (2004).

\bibitem{kikkawa:prl98}
Kikkawa, J.~M. and Awschalom, D.~D., ``Resonant spin amplification in n-type
  {GaAs},'' {\em Physical Review Letters}~{\bf 80},  4313 (1998).

\bibitem{heberle:prl01}
Sandhu, J.~S., Heberle, A.~P., Baumberg, J.~J., and Cleaver, J. R.~A.,
  ``Gateable suppression of spin relaxation in semiconductors,'' {\em Physical
  Review Letters}~{\bf 86}, 2150 (2001).

\bibitem{crooker:apl06-spin-relaxation-times-with-bias}
Furis, M., Smith, D.~L., Crooker, S.~A., and Reno, J.~L., ``Bias-dependent
  electron spin lifetimes in n-gaas and the role of donor impact ionization,''
  {\em Applied Physics Letters}~{\bf 89},  102102 (2006).

\bibitem{Crooker-prb97}
Crooker, S.~A., Awschalom, D.~D., Baumberg, J.~J., Flack, F., and Samarth, N.,
  ``Optical spin resonance and transverse spin relaxation in magnetic
  semiconductor quantum wells,'' {\em Physical Review B}~{\bf 56},
  7574 (1997).

\bibitem{ma:prl97}
Aeschlimann, M., Bauer, M., Pawlik, S., Weber, W., Burgermeister, R., Oberli,
  D., and Siegmann, H.~C., ``Ultrafast spin-dependent electron dynamics in fcc
  {Co},'' {\em Physical Review Letters}~{\bf 79},  5158 (1997).

\bibitem{wu:prb00:kinetics_qw}
Wu, M.~W. and Metiu, H., ``Kinetics of spin coherence of electrons in an
  undoped semiconductor quantum well,'' {\em Physical Review B}~{\bf 61},
  2945 (2000).

\bibitem{glazov-jetplett02}
Glazov, M.~M. and Ivchenko, E.~L., ``Precession spin relaxation mechanism
  caused by frequent electron-electron collisions,'' {\em Jetp Letters}~{\bf
  75},  403 (2002).

\bibitem{KraussPRL2008:holespin}
Krauss, M., Aeschlimann, M., and Schneider, H.~C., ``Ultrafast spin dynamics
  including spin-orbit interaction in semiconductors,'' {\em Physical Review
  Letters}~{\bf 100},  256601 (2008).

\bibitem{bap:76}
Bir, G.~L., Aronov, A.~G., and Pikus, G.~E., ``Spin relaxation of electrons due
  to scattering by holes,'' {\em Sov. Physics JETP}~{\bf 42},  705 (1976).

\bibitem{hcs-prb06:spin-relax-surface}
Schneider, H.~C., W{\"u}stenberg, J.~P., Andreyev, O., Hiebbner, K., Guo, L.,
  Lange, J., Schreiber, L., Beschoten, B., Bauer, M., and Aeschlimann, M.,
  ``Energy-resolved electron-spin dynamics at surfaces of p-doped {GaAs},''
  {\em Physical Review B}~{\bf 73},  081302(R) (2006).

\bibitem{maialle:prb96}
Maialle, M.~Z., ``Spin relaxation of electrons in p-doped quantum wells via the
  electron-hole exchange interaction,'' {\em Physical Review B}~{\bf 54},  1967
  (1996).

\bibitem{maialle:p-doped-holes:prb97}
Maialle, M.~Z. and Degani, M.~H., ``Electron-spin relaxation in p-type quantum
  wells via the electron-hole exchange interaction: The effects of the
  valence-band spin mixing and of an applied longitudinal electric field,''
  {\em Physical Review B}~{\bf 55},  13771 (1997).

\bibitem{song:prb02}
Song, P.~H. and Kim, K., ``Spin relaxation of conduction electrons in bulk
  {III-V} semiconductors,'' {\em Physical Review B}~{\bf 66}, 035207 (2002).

\bibitem{JiangWuPRB09:bulk-spin}
Jiang, J.~H. and Wu, M.~W., ``Electron-spin relaxation in bulk {III-V}
  semiconductors from a fully microscopic kinetic spin {Bloch} equation
  approach,'' {\em Physical Review B}~{\bf 79},  125206 (2009).

\bibitem{KraussPRB2010:bulk-GaAs}
Krauss, M., Schneider, H.~C., Bratschitsch, R., Chen, Z., and Cundiff, S.~T.,
  ``Ultrafast spin dynamics in optically excited bulk {GaAs} at low
  temperatures,'' {\em Physical Review B}~{\bf 81}, 035213 (2010).

\bibitem{oestreich-apl08:bulk-DP}
Oertel, S., H\"{u}bner, J., and Oestreich, M., ``High temperature electron spin
  relaxation in bulk gaas,'' {\em Applied Physics Letters}~{\bf 93},  132112
  (2008).

\bibitem{ZhouWuPRB2008}
Zhou, J. and Wu, M.~W., ``Spin relaxation due to the {Bir-Aronov-Pikus}
  mechanism in intrinsic and p-type {GaAs} quantum wells from a fully
  microscopic approach,'' {\em Physical Review B}~{\bf 77},  075318 (2008).

\bibitem{haug-koch}
Haug, H. and Koch, S.~W.,  {\em Quantum Theory of the Optical and Electronic
  Properties of Semiconductors}, World Scientific,
  Singapore, 4.~ed. (2004).

\bibitem{schaefer-book}
Sch{\"a}fer, W. and Wegener, M.,  {\em Semiconductor Optics and Transport
  Phenomena}, Springer (2002).

\bibitem{binder-koch}
Binder, R. and Koch, S.~W., ``Nonequilibrium semiconductor dynamics,'' {\em
  Prog. Quantum Electronics}~{\bf 19}(4/5),  307 (1995).

\bibitem{pikus-bir:71}
Pikus, G.~E. and Bir, G.~L., ``Exchange interaction in excitons in
  semiconductors,'' {\em Sov. Physics JETP}~{\bf 33},  208 (1971).

\bibitem{maialle:prb93:exciton-spin}
Maialle, M.~Z., Silva, E. A. D.~E., and Sham, L.~J., ``Exciton spin dynamics in
  quantum-wells,'' {\em Physical Review B}~{\bf 47},  15776 (1993).

\bibitem{cohen-tannoudji:qm}
Cohen-Tannoudji, C., Diu, B., and Lalo{\"e}, F.,  {\em Quantum
  Mechanics}, Wiley (1977).

\bibitem{collet:93}
Collet, J.~H., ``Screening and exchange in the theory of the femtosecond
  kinetics of the electron-hole plasma,'' {\em Physical Review B}~{\bf 47},
   10279 (1993).

\bibitem{hilton-tang}
Hilton, D.~J. and Tang, C.~L., ``Optical orientation and femtosecond relaxation
  of spin-polarized holes in {GaAs},'' {\em Physical Review Letters}~{\bf 89},  146601 (2002).

\end{thebibliography}

\end{document}